\documentclass[12pt,preprint]{aastex}
\usepackage{amsmath}	
\usepackage{epsfig}
\usepackage{natbib}                
\shorttitle{Interferometric analysis of three late-type stars}
\shortauthors{Boyajian \& McAlister et al.}
\begin{document}
\title{Angular Diameters of the G Subdwarf $\mu$ Cassiopeiae A \\
and the K Dwarfs $\sigma$ Draconis and HR~511 from \\
Interferometric Measurements with the CHARA Array}
\author{Tabetha S. Boyajian, Harold A. McAlister, \\ 
Ellyn K. Baines, Douglas R. Gies, Todd Henry, Wei-Chun Jao, David O'Brien, \\ 
Deepak Raghavan, Yamina Touhami}
\affil{Center for High Angular Resolution Astronomy, Georgia State University, P.O. Box 3969, Atlanta, GA 30302-3969}
\email{tabetha, hal, baines, gies, thenry, jao, obrien, \\
raghavan, yamina@chara.gsu.edu}
\author{Theo A. ten Brummelaar, Chris Farrington, P. J. Goldfinger,  \\ Laszlo Sturmann, Judit Sturmann, Nils H. Turner}
\affil{The CHARA Array, Mount Wilson Observatory, Mount Wilson, CA 91023}
\email{theo, farrington, pj, sturmann, judit, nils@chara-array.org}
\author{Stephen Ridgway}
\affil{National Optical Astronomy Observatory, P.O. Box 26732, Tucson, AZ 85726-6732}
\email{sridgway@noao.edu}
%
\begin{abstract}		
Using the longest baselines of the CHARA Array, we have measured the angular diameter of the G5~V subdwarf $\mu$~Cas~A, the first such determination for a halo population star.  We compare this result to new diameters for the higher metallicity K0~V stars, $\sigma$ Dra and HR~511, and find that the metal-poor star, $\mu$~Cas~A, has an effective temperature ($T_{\rm eff}=5297\pm32$~K), radius ($R=0.791\pm0.008 R_{\rm \odot}$), and absolute luminosity ($L=0.442\pm0.014 L_{\rm \odot}$) comparable to the other two stars with later spectral types.  We show that stellar models show a discrepancy in the predicted temperature and radius for $\mu$~Cas~A, and we discuss these results and how they provide a key to understanding the fundamental relationships for stars with low metallicity.  
\end{abstract}
%

\section{Introduction}

Direct measurements of stellar angular diameters offer a crucial means of providing accurate fundamental information for stars.  Advances in long baseline optical/infrared interferometry (LBOI) now enable us to probe the realm of cooler main-sequence stars to better define their characteristics. In their pioneering program at the Narrabri Intensity Interferometer, \citet{bro74} produced the first modern interferometric survey of stars by measuring the diameters of 32 bright stars in the spectral type range O5 to F8 with seven stars lying on the main sequence. The current generation of interferometers possesses sufficiently long baselines to expand the main sequence diameter sensitivity to include even later spectral types, as exemplified by \citet{lan01}, \citet{seg03}, and \citet{ber06} who determined diameters of K$-$M stars and \citet{bai08} who measured the radii of exoplanet host stars with types between F7 and K0.   

In this work, we focus primarily on the fundamental parameters of the well-known population~II star $\mu$~Cassiopeiae ($\mu$~Cas; HR~321; HD~6582; GJ~53~A), an astrometric binary with a period of $\sim 22$ years consisting of a G5 + M5 pair of main sequence stars with low metallicity (\citealt{dru95}, and references therein).  With the CHARA Array, we have measured the angular diameter of $\mu$~Cas~A to $<1\%$ accuracy, thereby yielding the effective temperature, linear radius, absolute luminosity and gravity (with accuracies of $0.6\%$, $1.0\%$, $3.2\%$, and $9.0\%$, respectively).  We compare these newly determined fundamental stellar parameters for $\mu$~Cas~A to those of two K0~V stars, HR~511 (HD~10780; GJ~75) and $\sigma$~Draconis ($\sigma$~Dra; HR~7462; HD 185144; GJ~764), which we also observed with the CHARA Array (\S4.1).  These fundamental parameters are then compared to model isochrones (\S4.2). 

%

\section{Interferometric Observations}

Observations were taken using the CHARA Array, located on Mount Wilson, CA, and remotely operated from the Georgia State University AROC\footnote{Arrington Remote Operations Center} facility in Atlanta, GA.  The data were acquired over several nights using a combination of the longest projected baselines (ranging from 230$-$320 m) and the CHARA Classic beam combiner in $K^{\prime}$-band \citep{ttb05}.  The data were collected in the usual calibrator-object-calibrator sequence (brackets), yielding a total of 26, 15, and 22 bracketed observations for $\mu$ Cas, $\sigma$~Dra, and HR~511, respectively.  

For both $\mu$~Cas~A and HR~511, we used the same calibrator star, HD~6210, which is a relatively close, unresolved, bright star with no known companions.  Under the same criteria, we selected HD~193664 as the calibrator star for $\sigma$~Dra.  For each star, a collection of magnitudes (Johnson $UBV$, \citealt{joh66}; Str\"{o}mgren $uvby$, \citealt{hau98}; {\it 2MASS} $JHK$, \citealt{skr06}) were transformed into calibrated flux measurements using the methods described in \citet{col96}, \citet{gra98}, and \citet{coh03}.   We then fit a model spectral energy distribution\footnote{The model fluxes were interpolated from the grid of models from R. L. Kurucz available at http://kurucz.cfa.harvard.edu/} (SED) to the observed flux calibrated photometry to determine the limb-darkened angular diameters $\theta_{\rm SED}$($T_{\rm eff}$, $\log g$) for these stars.  We find $\theta_{\rm SED}$(6100, 3.8)=$0.519 \pm 0.012$~mas for HD~6210 and $\theta_{\rm SED}$(6100, 4.5)=$0.494 \pm 0.019$~mas for HD~193664.  These angular diameters translate to absolute visibilities of 0.87 and 0.89 for the mean baselines used for the observations, or $\pm 0.8\%$ and $\pm 1.2\%$ errors, where these errors are propagated through to the final visibility measurements for our stars during the calibration process.  An additional independent source of error is the uncertainty in the effective wavelength of the observed spectral bandpass. As described by \citet{mca05}, the effective wavelength of the $K^{\prime}$ filter employed for these observations has been adjusted to incorporate estimates of the transmission and reflection efficiencies of the surfaces and mediums the light encounters on its way to the detector, as well as for the effective temperature of the star. This calculation yields an effective wavelength for these observations of $2.15 \pm 0.01 \mu$m, which leads to a contribution at the $0.4\%$ level to the angular diameter error budget.  Due to the fact that the flux distribution in the $K^{\prime}$-band for all of our stars is in the Rayleigh-Jeans tail, we find that there are no object-to-object differences in this calculation of effective wavelength due each star having a different effective temperature. 

%

\section{Data Reduction and Diameter Fits}

The data were reduced and calibrated using the standard data processing routines employed for CHARA Classic data (see \citealt{ttb05} and \citealt{mca05} for details).  For each calibrated observation, Table~1 lists the time of mid-exposure, the projected baseline $B$, the orientation of the baseline on the sky $\psi$, the visibility $V$, and the 1-$\sigma$ error to the visibility $\sigma (V)$.  

\placetable{tab1}	

We did not detect the secondary star in $\mu$~Cas as a separated fringe packet (SFP) in any of our observations (see \citealt{far06} for discussion on interferometric detections of SFP binaries).  However, for close binaries, the measured instrumental visibility is affected by the flux of two stars, so in addition to our analysis of $\mu$~Cas~A, we must account for incoherent light from the secondary star affecting our measurements.   By calculating the ephemeris positions of the binary at the time of our observations, we get the separation $\rho_{\rm AB}$ of the binary during each observation.  Although the most recent published orbital parameters are from \citet{dru95}, Gail Schaefer and collaborators (private communication) have provided us with their updated orbital elements for the binary based on Hubble Space Telescope observations taken every six months over the last decade.  We use these separations (ranging from $1.380-1.396$ arcseconds) in combination with $\Delta M_{K}=3.5$ for the binary \citep{mcc84} (assuming $K \approx K^{\prime}$) and seeing measurements at the time of each observation to calculate the amount of light the secondary contributes within our detector's field of view (details described in Appendix ~A).   Fortunately our correction factors to the visibilities of $\mu$~Cas~A are small ($0.4 - 1.4 \%$), so even high uncertainties in this correction factor have minimal impact on the final corrected measurement.

In order to obtain limb-darkening coefficients for our target stars, SED fits were made to estimate $T_{\rm eff}$ and $\log g$.  We used a bi-linear interpolation in the \citet{cla95} grid of linear limb-darkening coefficients in $K$-band ($\mu_{K}$) with our best fit SED parameters to get $\mu_{K}$ for each star.  Because limb darkening has minimal influence in the infrared (here, we also assume $K \approx K^{\prime}$), as well as minimal dependence on temperature, gravity, and abundance for these spectral types, we feel that this method is appropriate and at most  will contribute an additional one-tenth of one percent error
to our limb-darkened diameters.  We calculate the uniform-disk $\theta_{\rm UD}$ (Equation~1) and limb-darkened $\theta_{\rm LD}$ (Equation~2) angular diameters from the calibrated visibilities by $\chi^2$ minimization of the following relations \citep{bro74a}:
\begin{equation}
V  =  \frac{2 J_1(\rm x) }{\rm x},
\end{equation}
\begin{equation}
V = \left( {1-\mu_\lambda \over 2} + {\mu_\lambda \over 3} \right)^{-1} 
\times
\left[
(1-\mu_\lambda) {J_1(\rm x) \over \rm x} + \mu_\lambda {\left( \frac{\pi}{2} \right)^{1/2} \frac{J_{3/2}(\rm x)}{\rm x^{3/2}}} 
\right],
\end{equation}
and \begin{equation}
\rm x = \pi \emph{B} \theta \lambda^{-1},
\end{equation}
where $\emph{J}_n$ is the $n^{th}$-order Bessel function, and $\mu_{\lambda}$ is the linear limb darkening coefficient at the wavelength of observation.  In Equation~3, $\emph{B}$ is the projected baseline in the sky, $\theta$ is the UD angular diameter of the star when applied to Equation~1 and the LD angular diameter when used in Equation~2, and $\lambda$ is the central wavelength of the observational bandpass.  

The error to the diameter fit is based upon the values on either side of the minimum for which $\chi^2$ = $\chi^2_{\rm min}$ + 1 \citep{pre92,wal03}.  A summary of these results is presented in Table~2, and Figures~1$-$2 show the best fits to our calibrated visibilities along with the 1-$\sigma$ errors.  

\begin{figure}
\begin{center}
{\includegraphics[angle=90,height=12cm]{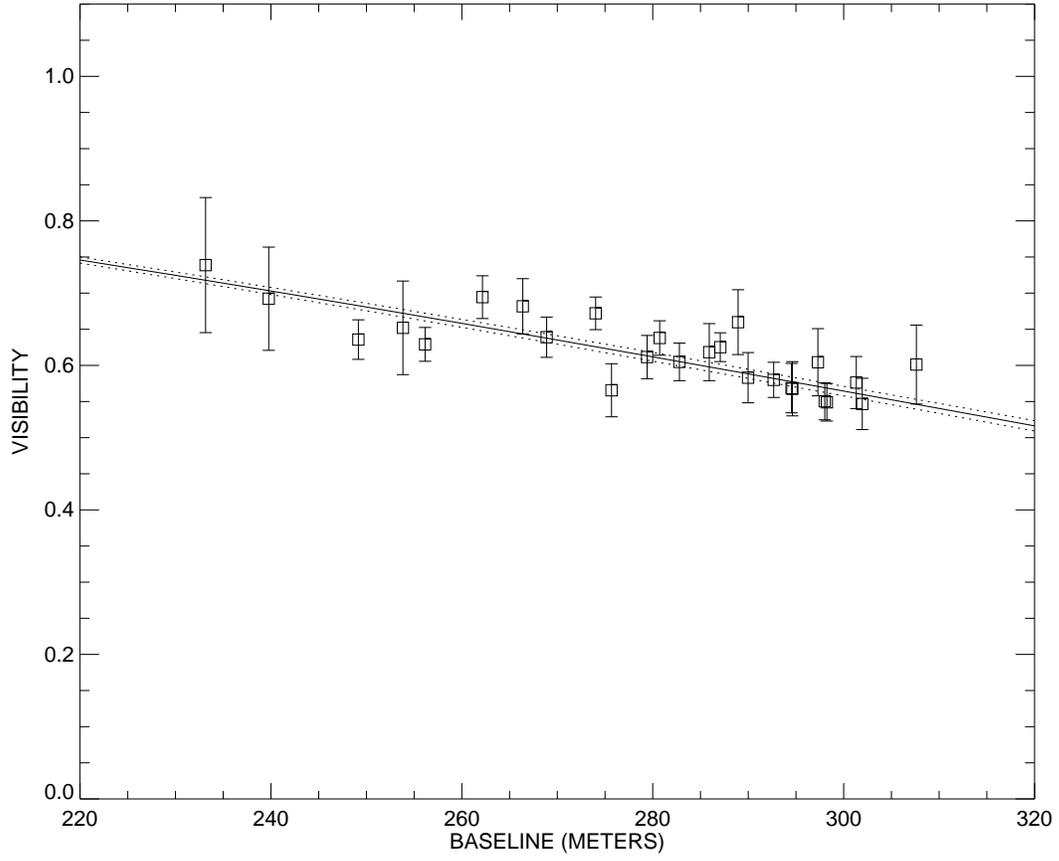}}
\end{center}
\caption{Limb-darkened angular diameter fit to $\mu$~Cas~A.}
\label{fig1}
\end{figure}

\begin{figure}
\begin{center}
{\includegraphics[angle=90,height=12cm]{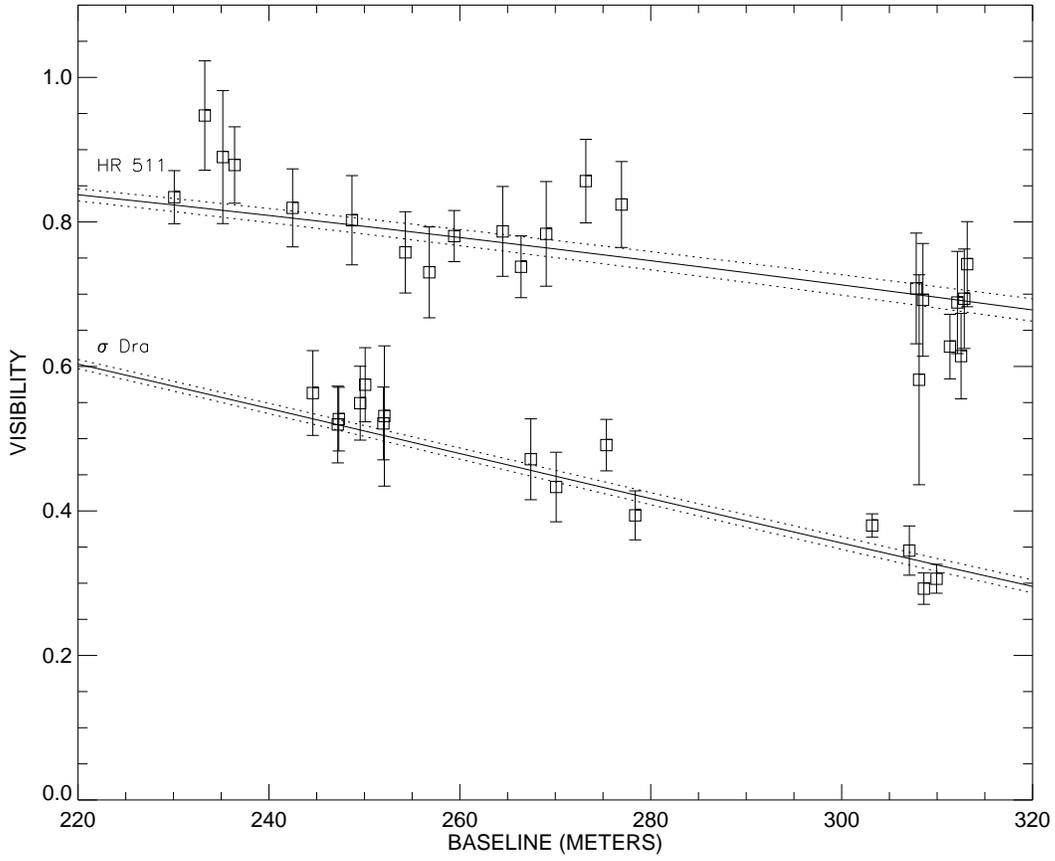}}
\end{center}
\caption{Limb-darkened angular diameter fit to $\sigma$~Dra ({\it bottom curve}) and HR~511 ({\it top curve}).}
\label{fig2}
\end{figure}

\placetable{tab2}	
\placefigure{fig1}     
\placefigure{fig2}     

%
\section{Discussion}

The linear radii, temperatures and absolute luminosities are calculated through fundamental relationships when the stellar distance, total flux received at Earth, and angular diameter are known.  The linear radius of each star can be directly determined by combining our measured angular diameter with the {\it Hipparcos} parallax. Next, the fundamental relation between a star's total flux $F_{\rm BOL}$ and angular diameter (Equation~4) is used to calculate the effective temperature $T_{\rm eff}$ and the absolute luminosity:  
\begin{equation}
F_{\rm BOL} =  \frac{1}{4} \theta_{\rm LD}^2 \sigma T_{\rm eff}^4,
\end{equation}
where $\sigma$ is the Stefan-Boltzmann constant.  For $\mu$~Cas~A, $\sigma$~Dra, and HR~511, we calculate radii, effective temperatures, and luminosities purely from direct measurements (Table~2). For these calculations, the $F_{\rm BOL}$ for $\mu$~Cas~A has been corrected for light contributed by the secondary by adopting the luminosity ratio of the two components from \citet{dru95}, effectively reducing its $F_{\rm BOL}$ by $1.3\%$.  

Table~2 lists our derived temperatures for $\mu$~Cas~A, $\sigma$~Dra, and HR~511 ($T_{\rm eff}=5297\pm32, 5299\pm32$, and $5350\pm76$~K, respectively).  Our temperatures agree well with the numerous indirect techniques used to estimate $T_{\rm eff}$ with spectroscopic or photometric relationships.  Temperatures of $\mu$~Cas~A derived using these methods range from 5091$-$5387~K (5143$-$5344~K for $\sigma$~Dra and 5250$-$5419~K for HR~511), and while the internal error is low in each reference, the apparent discrepancy among the various methods shows that there is some systematic offset for each temperature scale, as might be expected if atmospheric line opacities are not correctly represented in the models.

\subsection{Comparative Analysis to Observations of $\mu$~Cas~A, $\sigma$~Dra, and HR~511}		

It can be seen in Table~2 that the temperature, radius, and luminosity of $\mu$~Cas~A is quite similar to that of $\sigma$~Dra and HR~511 despite the large difference in spectral types and $B-V$ color indices associated with the classical characteristics of metal-poor stars.  These results support the conclusions in \citet{dru95}, where their model analysis predicts $\mu$~Cas~A to have the characteristic radius, temperature, and luminosity of a typical K0~V star.  In Figure~3, we compare our new linear radii versus $B-V$ color index to values measured from eclipsing binaries (EB's) and other LBOI measurements, as well as the position of the Sun and a theoretical ZAMS for solar metallicity stars.  The grayscale fill indicates metallicity estimates for the LBOI points, showing $\mu$~Cas~A is currently the lowest metallicity star observed in this region of the HR diagram.  The initial characteristics of evolution off the ZAMS is towards the upper-left region of the plot (larger and bluer), which is main reason for the dispersion of the stellar radii for stars in this region.  ZAMS lines for sub-solar metallicities lie below this line, and are shifted to bluer colors.  

\begin{figure}
\begin{center}
{\includegraphics[angle=90,height=12cm]{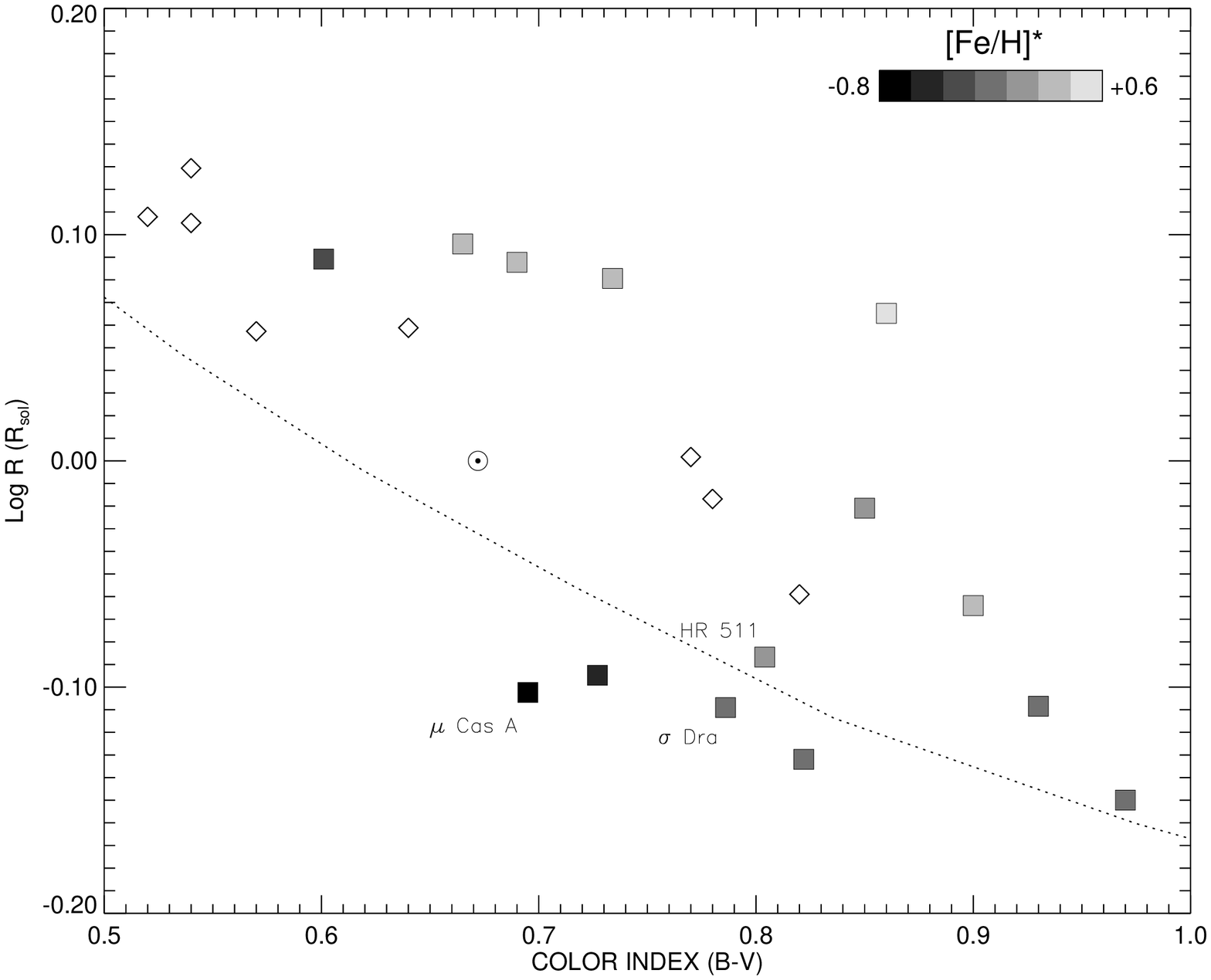}}
\end{center}
\caption{Plot of radius versus $B-V$ color index for G to mid-K type stars, including our results for $\mu$~Cas~A, $\sigma$~Dra, and HR~511.  Additional data plotted are from EB ({\it diamonds}, \citealt{and91}), LBOI ({\it squares}, \citealt{bai08}, \citealt{ker04}, and \citealt{lan01}), and the Sun ($\odot$).  The grayscale legend indicates the metallicity estimates for LBOI points from \citet{tay05} (EB metallicity estimates are unreliable due to their duplicity).  The {\it dotted} line represents a theoretical ZAMS line for solar metallicity (Y=0.275, Z=0.02) from the Dartmouth Stellar Evolution Models (\citealt{gue92,cha01}, available online at http://stellar.dartmouth.edu/). }
\label{fig3}
\end{figure}

\placefigure{fig3}     

\citet{dru95} determine the mass of $\mu$~Cas~A from the system's astrometric orbital solution.  \citet{leb99} update this mass utilizing the more accurate {\it Hipparcos} distance, yielding a mass of $0.757 \pm 0.060$ $M_{\rm \odot}$. We use this mass with our new radius, to derive a directly measured surface gravity of $\log g = 4.52 \pm 0.04$.  This value is comparable to the nominal values for solar metallicity ZAMS G5~V and K0~V stars being $\log g = 4.49$ \citep{cox00}.

We believe that the position of $\mu$~Cas~A on Figure~3 does not come from underestimated errors in our data, or in the archival data.  For instance, the uncertainty in the stellar radius can arise from the angular diameter we measure of the star (discussed in \S 3) and the {\it Hipparcos} parallax (Table~2).  Because of their nearness to the Sun, the parallaxes of all three of our stars are well determined by {\it Hipparcos}, and with the combined accuracy of our angular diameters, the uncertainty on these radii are all less than $3\%$.  The more pronounced discrepancy in the position of $\mu$~Cas~A on Figure~3 is the large offset in the $B-V$ color index for $\mu$~Cas~A with respect to the other two stars with the same effective temperatures, $\sigma$~Dra and HR~511.  However, according to the ranking system of \citet{nic78}, all three stars we analyze have the highest quality index of photometry, with a probable error in $B-V$ of $\pm$0.006.  In this catalog, the worst case scenario in photometry errors appears for characteristically dim stars with $V \gtrsim 10$, where the lowest rank quality index has an error in $B-V$ of $\pm$0.02, still not providing the desired effects to make the data agree within errors.  With regards to the binarity of $\mu$~Cas~A, the effect of the much cooler secondary star on the measured $B-V$ for the system as a whole would be less than one milli-magnitude \citep{cas07}, thus allowing us to ignore its contribution to these measurements as well.  In other words, the position of $\mu$~Cas~A on Figure~3 is simply a result of its lower metal abundance causing a reduction of opacity in its atmosphere, observationally making the star appear bluer in color than the other two stars with higher abundances with the same radius, effective temperature, and luminosity. 

We would like to make it clear that for $\mu$~Cas~A, a comparison of reduced opacities based solely on its iron abundance is a simplified approach, and complications arise in the determination of its true Helium abundance \citep{hay92} as well as enhanced $\alpha$-elements \citep{chi91}.  In this respect, reducing the Helium abundance, or increasing the $\alpha$-element abundance, mimics the effect of increasing the metal abundance on a star's effective temperature and luminosity.  Additionally, over timescales of 10 Gyr, microscopic diffusion must also be considered in abundance analyses of subdwarfs \citep{mor99}. Here, we do not wish to misrepresent the impact of these issues on various stellar parameters and modeling, but instead present a purely observational comparison to fundamentally observed properties of these three stars.  These topics will be discussed further in \S 4.2.

\subsection{Stellar Models}			

While we can achieve a substantial amount of information from eclipsing binaries such as mass and radius, there still exists great uncertainty in the effective temperatures and luminosities of these systems (for example, see the discussion in \S 3.4 and \S 3.5 in \citealt{and91}). On the other hand, while observing single stars with LBOI is quite effective in determining effective temperatures and luminosities of stars, it lacks the means of directly measuring stellar masses.  For $\mu$~Cas~A, the results of this work combined with our knowledge of the binary from previous orbital analysis provides us the best of both worlds.  Unfortunately, the current uncertainty in mass for $\mu$~Cas~A is $\sim10\%$, too great to produce useful information about the star when running model evolutionary tracks (see discussion below, and Figure~7).  However, our newly determined physical parameters of $\mu$~Cas~A provide us with a handy way to test the accuracy of stellar models for metal poor stars. 

To model $\mu$~Cas~A, $\sigma$~Dra, and HR~511, we use both the Yonsei-Yale (Y$^{2}$) stellar isochrones by \citet{yi01, kim02, yi03, dem04}, which apply the color table from \citet{lej98} and the Victoria-Regina (VR) stellar isochrones by \citet{van06} with $BVRI$ color-$T_{\rm eff}$ relations as described by \citet{van03}.  To run either of these model isochrones, input estimates are required for the abundance of iron [Fe/H] and $\alpha$-elements [$\alpha$/Fe], both of which contribute to the overall heavy-metal mass fraction $Z$.

The atmosphere of $\mu$~Cas is metal poor, and there are numerous abundance estimates ranging from [Fe/H]=$-$0.98 \citep{ful00} to [Fe/H]=$-$0.55 \citep{cle77}, with the most recent estimates favoring lower metallicity values.    Overall, this large range in metallicities suggests an error of $\Delta$[Fe/H]$\sim$0.2~dex. Systematic offsets aside, there exist a few additional variables which cause difficulties in determining accurate metallicity estimates for this star.  \citet{tor02} argue that abundance estimates for a binary are affected by the presence of the secondary in both photometric and spectroscopic measurement techniques.  However, \citet{wic74} measured the system's magnitude difference $\Delta$m$=5.5\pm0.7$ at $\lambda$=0.55$\mu$m, limiting the secondary's influence of these estimates to no more than $\Delta$[Fe/H]$\sim$0.05~dex, basically undetectable.  Secondly, the abundance analysis by \citet{the99} indicate that a careful non-LTE (NLTE) treatment is required when measuring stars with sub-solar abundances.  In the case of $\mu$~Cas~A, this correction factor is +0.14~dex, resulting in [Fe/H]$_{NLTE}=-0.56$ from their measurements.  Applying this correction factor brings the range of abundance estimates cited above to $-0.84<$[Fe/H]$<-0.41$.    

In this work, we use the averaged metallicity values from the \citet{tay05} catalog for all three stars (Table~2).  We caution the reader that this average value of [Fe/H] for $\mu$~Cas~A, corrected for NLTE effects described above, still lies below the value from \citet{the99} by about 0.12~dex; however, both of these estimates are within the range listed above.   \citet{leb99} show that indeed these corrections are needed to remove a large part of the discrepancy on model fits to match observations.  NLTE corrections for the iron abundance estimates of $\sigma$~Dra and HR~511 are not needed. 

$\sigma$~Dra and HR~511 show no sign of $\alpha$-enhanced elements with respect to the Sun (i.e., [$\alpha$/Fe]=0), which is not a surprise because they have near solar iron abundances \citep{mis04, sou05, ful00}.  However, these studies do detect the presence of $\alpha$-enhanced elements such as Ca, Mg, Si, and Ti in $\mu$~Cas~A, and we adopt an average value from these three sources to be [$\alpha$/Fe]$=0.36 \pm 0.06$.   

 To run models for each star, we round the average [Fe/H] value to the nearest [Fe/H] value in the VR models grids (Table~2), and adopt [$\alpha$/Fe]=0.3 for $\mu$~Cas~A, and [$\alpha$/Fe]=0.0 for $\sigma$~Dra and HR~511.  This approximation allows us to use identical input parameters in each of the models in order to compare the similarity of the models to each other (Figure~4).  To justify this approximation, we ran the Y$^{2}$ models (using the interpolating routine available) for both the exact and rounded input parameters for $\mu$~Cas~A and we were not able to see any substantial differences in comparing the two.  

We show our results compared to the Y$^{2}$ (left column) and VR (right column) stellar isochrones in Figures~4$-$6 in both the temperature and color dependent planes. The sensitivity to age in this region is minimal, but for reference, we plotted 1, 5, and 10~Gyr isochrones for each model, as well as the positions of $\mu$~Cas~A, $\sigma$~Dra and HR~511.  When comparing the model isochrones for $\mu$~Cas~A in Figure~4, no significant differences are seen between the Y$^{2}$ and VR models.  However, for both models these results show that there exist discrepancies to observations in the $T_{\rm eff}$ plane.  Both of the models overpredict the temperature for $\mu$~Cas~A for a given luminosity and radius.  On the other hand, on the color dependent plane,  the models appear to do an adequate job fitting the observations in terms of luminosity (for a typical age of a halo star of $\sim$10~Gyr), but an offset is still seen in the model radii versus color index.  In regards to model isochrones run for $\sigma$~Dra and HR~511 (Figures 5, 6), both of which have abundances more similar to the Sun, we find that the models and our observations agree quite well. 

\begin{figure}
\begin{center}
{\includegraphics[angle=0, height=16cm]{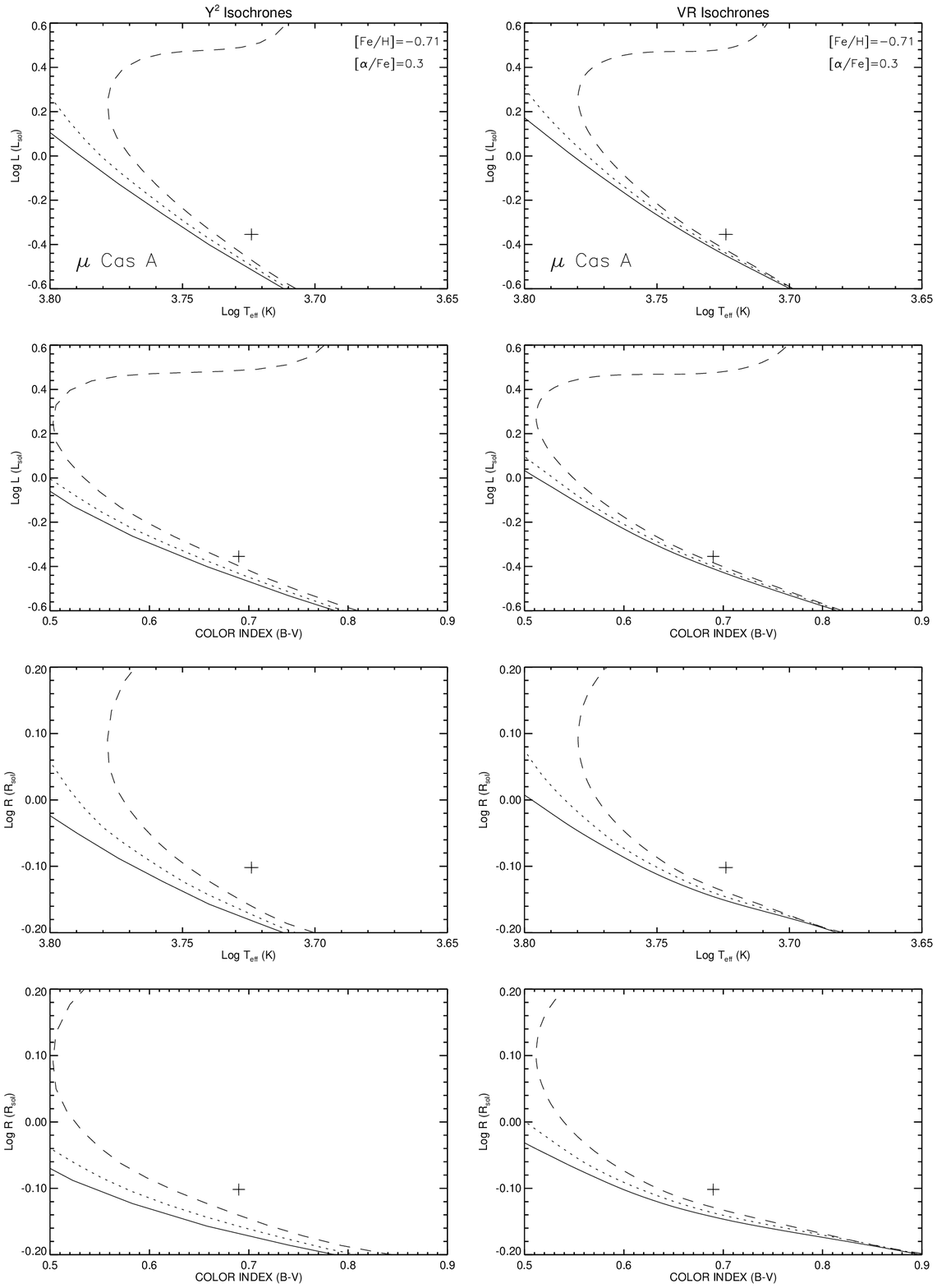}}
\end{center}
\caption{$\mu$~Cas~A data (along with 1-$\sigma$ errors) plotted against Y$^{2}$ and Victoria-Regina (VR) isochrones ([$\alpha$/Fe]=0.3, [Fe/H]=$-$0.71) for 1, 5, and 10~Gyr ({\it solid, dotted}, and {\it dashed} lines, respectively).}
\label{fig4}
\end{figure}

\begin{figure}
\begin{center}
{\includegraphics[angle=0, height=16cm]{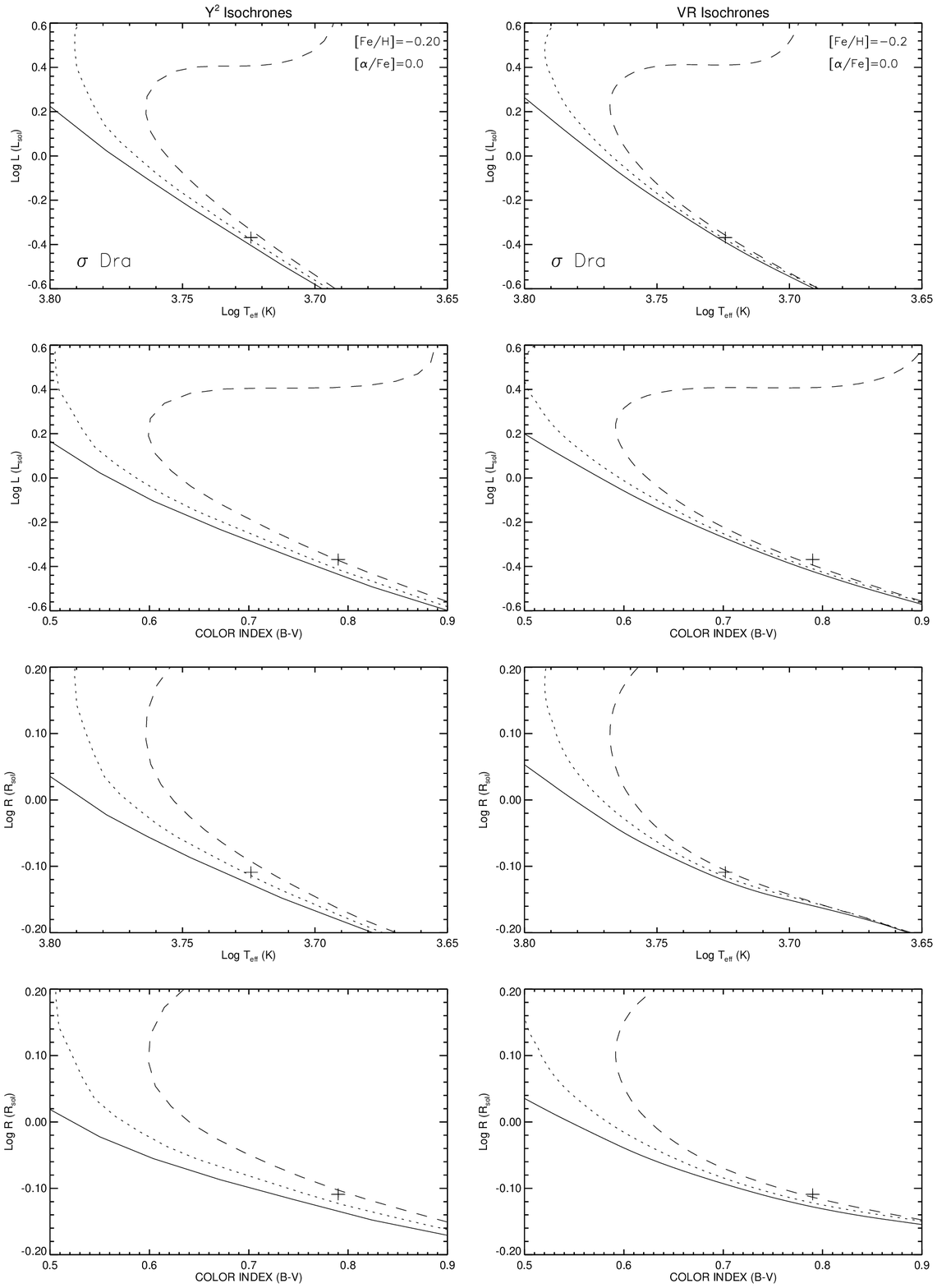}}
\end{center}
\caption{$\sigma$~Dra data (along with 1-$\sigma$ errors) plotted against Y$^{2}$ and Victoria-Regina (VR) isochrones ([$\alpha$/Fe]=0.0, [Fe/H]=$-$0.20) for 1, 5, and 10~Gyr ({\it solid, dotted}, and {\it dashed} lines, respectively).}
\label{fig5}
\end{figure}

\begin{figure}
\begin{center}
{\includegraphics[angle=0, height=16cm]{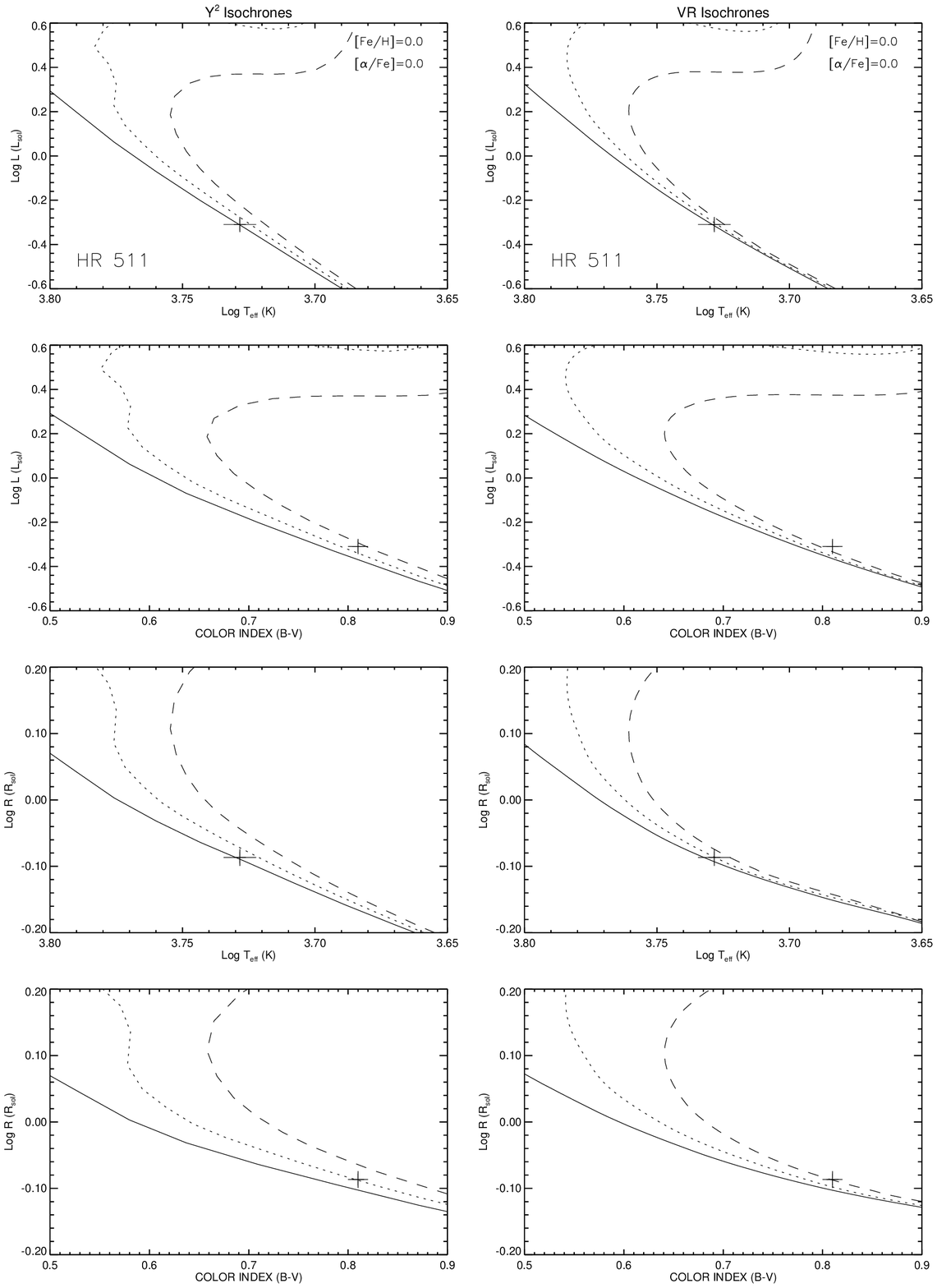}}
\end{center}
\caption{HR~511 data (along with 1-$\sigma$ errors) plotted against Y$^{2}$ and Victoria-Regina (VR) isochrones ([$\alpha$/Fe]=0.0, [Fe/H]=0.00) for 1, 5, and 10~Gyr ({\it solid, dotted}, and {\it dashed} lines, respectively).}
\label{fig6}
\end{figure}

\placefigure{fig4}     
\placefigure{fig5}     
\placefigure{fig6}     

It is apparent in Figure~4 that although both of the models are fairly consistent with each other, the methods used to transform $B-V$ color index to $T_{\rm eff}$ for metal poor stars is not calibrated correctly.  Likewise, as described by \citet{pop97} as being a ``serious dilemma,'' several recent works (utilizing all the current measurements of stellar radii measured) show that models are infamous in predicting temperatures that are too high and radii that are too small, while still being able to correctly reproduce the stellar luminosity (e.g. \citealt{mor08, rib07, lop07}).  Explanations for these discrepancies are understood to be a consequence of the stellar metallicity, magnetic activity, and/or duplicity.  

\placefigure{fig5}     

In Figure~7, we show our observations dependent on stellar mass compared to the model Y$^{2}$ isochrones $\mu$~Cas~A (VR models are not shown for clarity, but display approximately the same relations).  Here it is clear that the current errors in the measured mass for $\mu$~Cas~A is not sufficiently constrained to conclude anything useful from the models.  Fundamental properties of the secondary star are also an important constraint within these parameters, especially in the respects of co-evolution of the binary, but unfortunately both the Y$^{2}$ and VR models do not extend to masses low enough to test these issues.

%
%
\section{Conclusion}

In this first direct measurement of the diameter of a subdwarf, we find that although $\mu$~Cas~A is classified as a G5~V star, its sub-solar abundance leads it to resemble a K0~V star in terms of temperature, radius, and luminosity, whereas its surface gravity reflects the value for G5-K0 ZAMS stars with solar abundances.  We find that while the both Y$^{2}$ and VR isochrones agree with our observations of $\sigma$~Dra and HR~511, a discrepancy is seen in temperature and radius when comparing these models to our observations of $\mu$~Cas~A. 

We are currently working on modeling this star and other subdwarfs with hopes to better constrain stellar ages and composition.  Future plans to observe more stars of similar spectral types to determine angular diameters for main sequence stars are planned by TSB.  This work will accurately determine the fundamental characteristics of temperature, radius, and absolute luminosity of a large sample of stars and thereby contribute to a broad range of astronomical interests.

%
%
\acknowledgements

We would like to thank Gail Schaefer for sharing her preliminary results with us for the orbit of $\mu$~Cas.  We would also like to thank Gerard T. van Belle and David H. Berger for advice on the preparation of the project, as well as to Andrew Boden for advice on analyzing the data.  The CHARA Array is funded by the National Science Foundation through NSF grant AST-0606958 and by Georgia State University through the College of Arts and Sciences. This research has made use of the SIMBAD literature database, operated at CDS, Strasbourg, France, and of NASA's Astrophysics Data System. This publication makes use of data products from the Two Micron All Sky Survey, which is a joint project of the University of Massachusetts and the Infrared Processing and Analysis Center/California Institute of Technology, funded by the National Aeronautics and Space Administration and the National Science Foundation.

%
%
\clearpage
\appendix
\section{Appendix}
We translate our measurement of the Fried parameter $r_{0}$ into the astronomical seeing disk $\theta_{\rm Seeing}$ by 
\begin{equation}
r_{0} = 1.009 D \left(\frac{\lambda}{\theta_{\rm Seeing}D}\right)^{6/5}
\end{equation}
where $D$ is telescope aperture size, and $\lambda$ is wavelength of observation \citep{ttb93}.  To first order, an adequate representation of the intensity distribution of light from a star is a Gaussian \citep{kin71,rac96}, where $\theta_{\rm Seeing}$ is modeled as the full with at half maximum of the Gaussian .  Thus, we can write the normalized intensity distribution of light for a star as
\begin{equation}
I(x,x_0,y,y_0) = \frac{1}{2 \pi \sigma^{2}} \exp\left[{-\frac{1}{2 \sigma^2} \left[(x-x_0)^2+(y-y_0)^2\right]}\right] 
\end{equation}
where $\sigma \equiv 2.355^{-1}~\theta_{\rm Seeing}$, and the coordinates ($x_0$,$y_0$) determine the central position of  the star on the chip.  Assuming the primary star is at the center of our $2\times2$ pixel array ($0,0$) and the secondary is offset by its separation in arcseconds ($0,\rho_{\rm AB}$), we then have the amount of light contributed by each star:
\begin{equation}
I_A = Q \int^1_{-1} \int^1_{-1} I(x,0,y,0) \,dx\,dy 
\end{equation}
and
\begin{equation}
I_B = \int^1_{-1} \int^1_{-1} I(x,0,y,\rho_{\rm AB}) \,dx\,dy
\end{equation}
with $Q$ being the intensity ratio of the two stars, $Q=10^{\Delta{\rm M_K}/2.5}$.  Hence, the conversion of the measured visibility $V$ to the true visibility for the primary star $V_{A}$ is 
\begin{equation}
V_A = V (1+I_B/I_A).
\end{equation}

%
\clearpage
\bibliographystyle{apj}             
\bibliography{apj-jour,paper}       
\clearpage
%
%
\begin{deluxetable}{lccccc}
\tablewidth{0pc} \tablecaption{Interferometric Measurements}
\tablehead{
 \colhead{Star} & \colhead{JD} &  \colhead{$B$} & \colhead{$\psi$}      &\colhead{$V$\tablenotemark{a}} & \colhead{$\sigma (V)$} \\
 \colhead{} & \colhead{($-$2,400,000)}   & \colhead{(m)} & \colhead{($^{\circ}$)}     & \colhead{ }          & \colhead{ } }
 \startdata
$\mu$~Cas~A	 &	54282.917	&	233.2	&	135.0	&	0.739	&	0.093	\\
$\mu$~Cas~A	 &	54282.929	&	239.8	&	130.0	&	0.692	&	0.071	\\
$\mu$~Cas~A	 &	54282.954	&	253.8	&	120.4	&	0.652	&	0.065	\\
$\mu$~Cas~A	 &	54298.915	&	266.4	&	234.3	&	0.682	&	0.038	\\
$\mu$~Cas~A	 &	54298.929	&	274.0	&	231.4	&	0.672	&	0.023	\\
$\mu$~Cas~A	 &	54298.942	&	280.7	&	228.6	&	0.638	&	0.024	\\
$\mu$~Cas~A	 &	54298.957	&	287.1	&	225.6	&	0.625	&	0.020	\\
$\mu$~Cas~A	 &	54298.971	&	292.7	&	222.7	&	0.580	&	0.024	\\
$\mu$~Cas~A	 &	54298.986	&	298.0	&	219.4	&	0.550	&	0.026	\\
$\mu$~Cas~A	 &	54299.885	&	249.2	&	239.9	&	0.636	&	0.027	\\
$\mu$~Cas~A	 &	54299.896	&	256.2	&	237.8	&	0.629	&	0.023	\\
$\mu$~Cas~A	 &	54299.905	&	262.2	&	235.8	&	0.694	&	0.030	\\
$\mu$~Cas~A	 &	54299.917	&	268.9	&	233.4	&	0.639	&	0.028	\\
$\mu$~Cas~A	 &	54299.961	&	290.0	&	224.1	&	0.583	&	0.035	\\
$\mu$~Cas~A	 &	54299.973	&	294.6	&	221.5	&	0.568	&	0.038	\\
$\mu$~Cas~A	 &	54299.984	&	298.2	&	219.2	&	0.549	&	0.026	\\
$\mu$~Cas~A	 &	54299.996	&	301.9	&	216.6	&	0.547	&	0.035	\\
$\mu$~Cas~A	 &	54351.787	&	275.7	&	219.2	&	0.566	&	0.037	\\
$\mu$~Cas~A	 &	54351.795	&	279.4	&	220.8	&	0.612	&	0.030	\\
$\mu$~Cas~A	 &	54351.802	&	282.8	&	222.3	&	0.605	&	0.026	\\
$\mu$~Cas~A	 &	54351.809	&	285.9	&	223.8	&	0.618	&	0.040	\\
$\mu$~Cas~A	 &	54351.816	&	288.9	&	225.3	&	0.660	&	0.045	\\
$\mu$~Cas~A	 &	54351.831	&	294.5	&	228.4	&	0.569	&	0.034	\\
$\mu$~Cas~A	 &	54351.839	&	297.3	&	230.2	&	0.604	&	0.047	\\
$\mu$~Cas~A	 &	54351.851	&	301.3	&	232.9	&	0.576	&	0.036	\\
$\mu$~Cas~A	 &	54351.875	&	307.6	&	238.3	&	0.601	&	0.055	\\	\\
$\sigma$~Dra	 &	54244.974	&	252.1	&	134.9	&	0.531	&	0.097	\\
$\sigma$~Dra	 &	54244.984	&	250.1	&	131.7	&	0.575	&	0.051	\\
$\sigma$~Dra	 &	54244.997	&	247.3	&	127.8	&	0.527	&	0.044	\\
$\sigma$~Dra	 &	54245.971	&	252.0	&	134.7	&	0.521	&	0.050	\\
$\sigma$~Dra	 &	54245.984	&	249.6	&	131.0	&	0.549	&	0.051	\\
$\sigma$~Dra	 &	54245.995	&	247.2	&	127.7	&	0.520	&	0.053	\\
$\sigma$~Dra	 &	54246.007	&	244.6	&	124.3	&	0.563	&	0.059	\\
$\sigma$~Dra	 &	54279.838	&	303.2	&	268.9	&	0.380	&	0.016	\\
$\sigma$~Dra	 &	54280.715	&	275.4	&	131.8	&	0.491	&	0.036	\\
$\sigma$~Dra	 &	54280.860	&	307.1	&	260.5	&	0.345	&	0.034	\\
$\sigma$~Dra	 &	54280.872	&	308.6	&	256.6	&	0.292	&	0.022	\\
$\sigma$~Dra	 &	54280.884	&	309.9	&	252.5	&	0.306	&	0.020	\\
$\sigma$~Dra	 &	54281.725	&	278.4	&	127.1	&	0.394	&	0.034	\\
$\sigma$~Dra	 &	54282.675	&	267.4	&	145.5	&	0.472	&	0.056	\\
$\sigma$~Dra	 &	54282.687	&	270.1	&	140.5	&	0.433	&	0.048	\\	\\
HR~511	 &	52922.857	&	235.2	&	137.4	&	0.890	&	0.092	\\
HR~511	 &	52922.867	&	233.3	&	140.4	&	0.947	&	0.076	\\
HR~511	 &	54280.952	&	256.8	&	138.4	&	0.730	&	0.063	\\
HR~511	 &	54280.979	&	266.4	&	127.5	&	0.738	&	0.043	\\
HR~511	 &	54301.903	&	230.1	&	248.9	&	0.834	&	0.037	\\
HR~511	 &	54301.913	&	236.4	&	246.2	&	0.879	&	0.053	\\
HR~511	 &	54301.924	&	242.5	&	243.4	&	0.819	&	0.054	\\
HR~511	 &	54301.935	&	248.7	&	240.5	&	0.802	&	0.062	\\
HR~511	 &	54301.946	&	254.3	&	237.7	&	0.758	&	0.056	\\
HR~511	 &	54301.957	&	259.4	&	235.0	&	0.780	&	0.035	\\
HR~511	 &	54301.968	&	264.5	&	232.2	&	0.787	&	0.062	\\
HR~511	 &	54301.979	&	269.0	&	229.5	&	0.783	&	0.072	\\
HR~511	 &	54301.989	&	273.2	&	226.8	&	0.856	&	0.058	\\
HR~511	 &	54302.000	&	276.9	&	224.2	&	0.824	&	0.059	\\
HR~511	 &	54383.935	&	313.2	&	220.9	&	0.742	&	0.059	\\
HR~511	 &	54383.943	&	312.8	&	223.7	&	0.694	&	0.069	\\
HR~511	 &	54383.950	&	312.5	&	226.2	&	0.614	&	0.059	\\
HR~511	 &	54383.958	&	312.1	&	228.7	&	0.688	&	0.071	\\
HR~511	 &	54383.971	&	311.3	&	233.2	&	0.627	&	0.045	\\
HR~511	 &	54384.017	&	308.5	&	249.0	&	0.692	&	0.078	\\
HR~511	 &	54384.025	&	308.1	&	251.6	&	0.582	&	0.145	\\
HR~511	 &	54384.031	&	307.8	&	253.9	&	0.708	&	0.077	\\
\enddata
\tablenotetext{a}{Corrected for light from secondary for $\mu$~Cas~A, see \S 3}
\end{deluxetable}
\clearpage
%
%
\begin{deluxetable}{lccc}
\tabletypesize{\scriptsize}
\tablewidth{0pc}
\tablecaption{Stellar Parameters}
\tablehead{
\colhead{Element} & 
\colhead{$\mu$~Cas~A}  &
\colhead{$\sigma$~Dra} &
\colhead{HR~511}  
} 
\startdata
Spectral Type		\dotfill	& G5~Vp				& K0~V				& K0~V				\\	
$V$ mag			\dotfill	& \phn5.17			& \phn4.70			& \phn5.63			\\
$B-V$			\dotfill	& \phn0.69			& \phn0.79			& \phn0.81			\\
$\pi_{hip}$ (mas)       \dotfill	& $132.42 \pm 0.60$\phn\phn	& $173.40 \pm 0.46$\phn\phn	& $100.24 \pm 0.68$\phn\phn	\\ 
$\theta_{\rm UD}$ (mas)	\dotfill	& $0.951 \pm 0.009$		& $1.224 \pm 0.011$		& $0.747 \pm 0.021$		\\
Reduced $\chi^2_{\rm UD}$\dotfill	& \phn$0.96$			& \phn$1.00$			& \phn$0.78$			\\
$\theta_{\rm LD}$ (mas)	\dotfill	& $0.973 \pm 0.009$		& $1.254 \pm 0.012$		& $0.763 \pm 0.021$		\\
Reduced $\chi^2_{\rm LD}$\dotfill	& \phn$0.96$			& \phn$1.01$			& \phn$0.79$			\\
Radius ($R_{\rm \odot}$)\dotfill        & $0.791 \pm 0.008$		& $0.778 \pm 0.008$		& $0.819 \pm 0.024$		\\
$F_{\rm BOL}$ (erg s$^{-1}$ cm$^{-2}$)\tablenotemark{a}\dotfill & $2.482 E-7$\tablenotemark{b}& $4.130 E-7\tablenotemark{c}$			& $1.588 E-7$\tablenotemark{d}			\\
$[$Fe/H$]$\tablenotemark{e}	\dotfill	& \phd$-0.682$\tablenotemark{f}($-$0.71)	& \phd$-0.199$\tablenotemark{g}($-$0.20)	& \phn\phn$0.005$\tablenotemark{g}(0.00)		\\ 
$T_{\rm eff}$ (K)	\dotfill	& $5297 \pm 32$\phn\phn		& $5299 \pm 32$\phn\phn		& $5350 \pm 76$\phn\phn			\\
Luminosity ($L_{\rm \odot}$) \dotfill   & $0.442 \pm 0.014$		& $0.428 \pm 0.013$		& $0.49 \pm 0.04$		\\
$\log g$ (cgs) \dotfill	&	$4.52 \pm 0.04$	&	\nodata	&	\nodata	\\
\enddata
\tablenotetext{a}{Adopted $1.5\%$ error}
\tablenotetext{b}{Average from \citet{bla98} and \citet{alo96}}
\tablenotetext{c}{Average from \citet{bel89} and \citet{alo96}}
\tablenotetext{d}{\citet{alo95}}
\tablenotetext{e}{Number in parenthesis is metallicity value used in models}
\tablenotetext{f}{\citet{tay05} +0.14~dex NLTE correction from \citet{the99}}
\tablenotetext{g}{\citet{tay05}}
\end{deluxetable}
\clearpage
%
\input{epsf}
\clearpage
%
\input{epsf}
\clearpage
\clearpage
\input{epsf}
%
\clearpage
\input{epsf}
%
\clearpage
\input{epsf}
%
\clearpage
\input{epsf}
%
\clearpage
\input{epsf}
\begin{figure}
\begin{center}
{\includegraphics[angle=90,height=12cm]{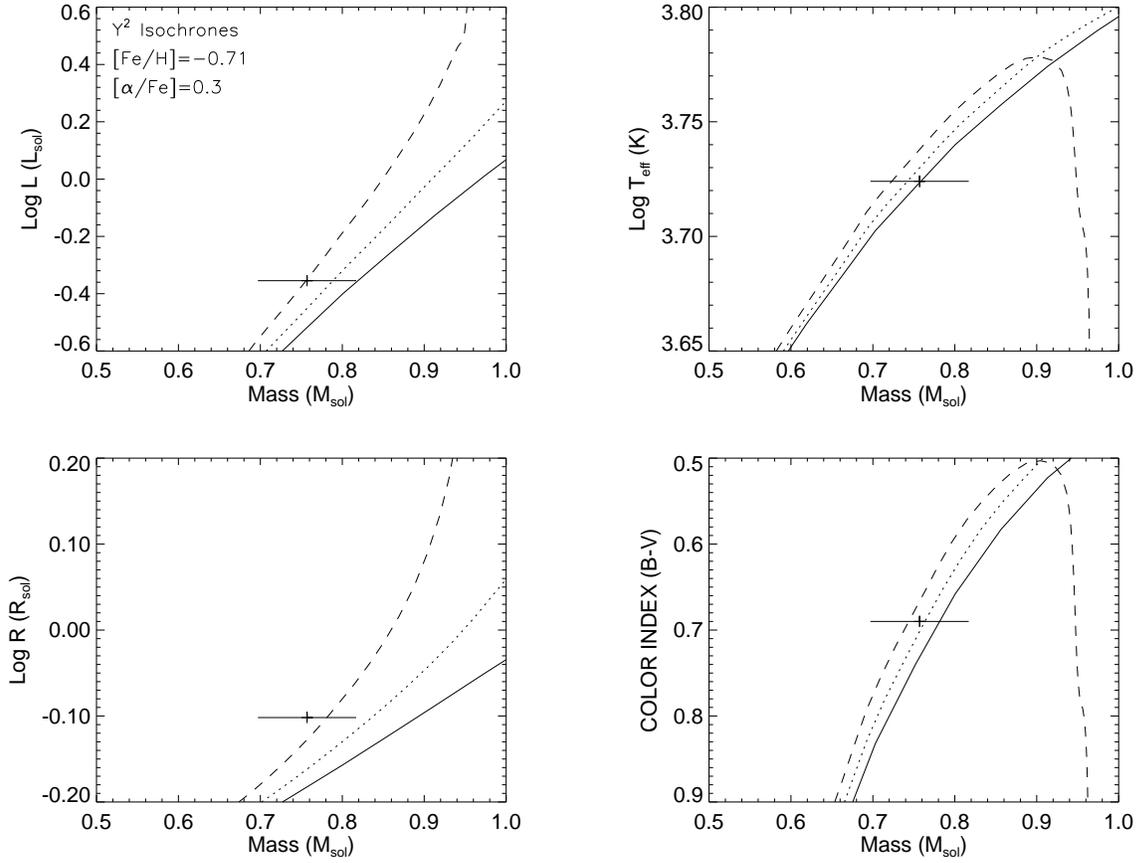}}
\end{center}
\caption{Mass relationships (along with 1-$\sigma$ errors) for $\mu$~Cas~A compared to Y$^{2}$ isochrones for 1, 5, and 10~Gyr ({\it solid, dotted}, and {\it dashed} lines, respectively).}
\label{fig7}
\end{figure}
%


\end{document}